\documentclass[12pt]{iopart}

\usepackage{epsfig}

\newcommand{\beq}{\begin{equation}}
\newcommand{\eeq}{\end{equation}}
\newcommand{\beqa}{\begin{eqnarray}}
\newcommand{\eeqa}{\end{eqnarray}}
\newcommand{\om}{\Omega_m}
\newcommand{\omb}{\Omega_b}
\newcommand{\fg}{f_{\rm gas}}
\newcommand{\zm}{z_{\rm max}}

\def\ga{\mathrel{\mathpalette\fun >}}
\def\fun#1#2{\lower3.6pt\vbox{\baselineskip0pt\lineskip.9pt
  \ialign{$\mathsurround=0pt#1\hfil##\hfil$\crcr#2\crcr\sim\crcr}}}

\begin{document} 

\title{Cosmology with X-ray Cluster Baryons} 
\author{Eric V.\ Linder} 
\address{Berkeley Lab, University of California, Berkeley, CA 94720 USA} 
\ead{evlinder@lbl.gov}

\begin{abstract} 
X-ray cluster measurements interpreted with a universal baryon/gas mass 
fraction can theoretically serve as a cosmological distance probe.  We 
examine issues of cosmological sensitivity for current (e.g.\ Chandra 
X-ray Observatory, XMM-Newton) and next generation (e.g.\ Con-X, XEUS) 
observations, 
along with systematic uncertainties and biases.  To give competitive 
next generation constraints on dark energy, we find that systematics 
will need to be controlled to better than 1\% and any evolution 
in $f_{\rm gas}$ (and other cluster gas properties) must be calibrated 
so the residual uncertainty is weaker than $(1+z)^{0.03}$. 

\end{abstract} 




\section{Introduction} \label{sec:intro}

An increasing number of cosmological methods have been suggested to 
explore the cosmological model and the accelerating universe.  This 
diversity offers hope in understanding the nature of our universe, if 
the probes are robust and clear in interpretation.  Astrophysical 
systematic uncertainties of a technique and its observations 
limit the cosmological leverage, despite statistical precision. 
Here we examine the level of control of systematics required for accurate 
cosmology estimation from the X-ray cluster distance method, sometimes 
called the baryon or gas mass fraction method, that assumes a universal 
baryon/gas mass fraction. 

X-ray cluster observations of fewer than 50 clusters by the Chandra 
X-ray Observatory \cite{chandra} and XMM-Newton \cite{xmm} have already 
been claimed to yield cosmological 
constraints \cite{ettori,allen04,weller,rapetti}.  For the future, there 
are prospects of more and deeper clusters, for example from Constellation-X 
\cite{conx} and XEUS \cite{xeus}.  We investigate the possible constraints, 
their cosmological degeneracies, and complementarity with other probes, 
giving special attention to the effect of hypothetical systematic floors 
in the achievable accuracy.  While the X-ray cluster baryon technique 
requires understanding the influence of cosmology and astrophysics on 
the ingredients of cosmic geometry, mass distributions, and hydrodynamical 
gas properties, it also has rich observational data. 

A number of recent papers have raised issues concerning the central 
assumption of universality of the cluster gas mass fraction, from 
observational, simulation, and theory points of view, e.g.\ 
\cite{vikhlinin,ettorifgas,hallman,loeb,nagai,ferramacho,gastaldello}.  
Given this uncertainty (which this paper does not try to resolve), it is 
useful to investigate how an overall uncertainty in the technique, 
arising from whatever source, affects the cosmological conclusions. 
We emphasize that this paper presents purely an investigation of 
cosmological sensitivity to the innate parameter degeneracies and 
end-of-pipeline level of systematic uncertainties; we do not make 
claims for the error contribution from specific measurements or what 
level of end systematics is achievable. 

In \S\ref{sec:method} 
we examine the cosmological sensitivity of the constant gas mass 
fraction technique based 
on parameter degeneracies and survey depth. Biases in the 
cosmological results arising from possible systematics are calculated 
in \S\ref{sec:bias}.  A brief review of the gas mass fraction technique, 
noting assumptions and possible areas for systematic uncertainties, 
is presented in the Appendix, along with a comment on bias from 
non-Gaussian errors in the translation to a distance measure.

\section{X-ray Cluster Gas as Distance Indicator} \label{sec:method} 

The X-ray flux from a galaxy cluster is related to the cosmological 
parameters we seek to estimate by 
\beq 
F_X=\tilde A\, d_a^{-3}\fg^2=A\,d_a^{-3}(\omb/\om)^2, \label{eq:fxd}
\eeq 
(see the Appendix for a review of the derivation) where $d_a$ is the 
angular diameter distance to the cluster, 
$\fg$ is the cluster gas mass fraction and $(\omb/\om)$ the universal 
baryon mass fraction, where $\omb$ is the baryon density in units of 
the critical density and $\om$ the matter density in units of the 
critical density.  The quantities $\tilde A$, $A$ should be independent 
of redshift and cosmology. 

In practice, one assumes $\fg$ should be constant (basically the universal 
value) and interprets any deviation from constancy as a deviation from 
the fiducial cosmology, adjusting the cosmology until $\fg(z)$ appears 
constant.  However, if $\fg$ is not truly constant over the redshift 
range of the survey then one will be led to an improper, biased 
cosmology.  Several papers recently 
\cite{vikhlinin,ettorifgas,hallman,loeb,nagai,ferramacho,gastaldello} 
indicate that the question of constancy is still open.  Without weighing in 
on this controversy, or the astrophysical assumptions and resulting 
individual systematic uncertainties that go into deriving Eq.~(\ref{eq:fxd}) 
(see the Appendix for a quick review), let us look at the big picture 
and ask at what level we must control the overall systematic uncertainties 
embodied in the proportionality factor $A$ (which includes both $\fg$ and 
other possible uncertainties).  Again, any redshift dependence of $A$ 
will bias the cosmology. 

First, we investigate the cosmological parameter sensitivities and 
degeneracies.  As the central cosmological quantity, we take the 
combination $X\equiv d_a^{-3}(\omb/\om)^2$ from 
Eq.~(\ref{eq:fxd}). 
In this section we assume that the baryon density is known, an 
optimistic assumption since Big Bang nucleosynthesis and cosmic 
microwave background measurements determine the quantity $\omb h^2$ 
rather than $\omb$, where $h$ is the reduced Hubble constant.  In 
\S\ref{sec:bias} we return to $\omb$ and other neglected ``constants'' 
of proportionality.  In the Appendix we discuss cautions regarding 
cosmological analysis by transforming $X$ directly to a distance quantity 
or $\fg$. 

Considering measurements $X(z)$ at various redshifts $z$, we 
calculate the cosmological sensitivity $\partial X/\partial p$ and 
joint likelihood contours by Fisher analysis (approximating the 
likelihoods as Gaussian) for the parameter set 
$p=\{\om,w_0,w_a\}$,  with a fiducial flat 
$\Lambda CDM$, $\om=0.28$ model.  The dark energy equation of state 
is modeled by $w(z)=w_0+w_a(1-a)$, where $a=1/(1+z)$. 

\begin{figure}[!hbt]
\begin{center} 
\psfig{file=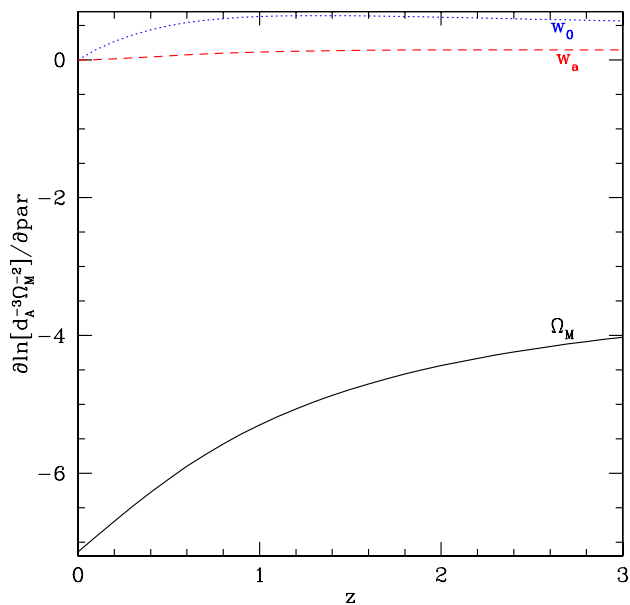,width=3.4in} 
\caption{The sensitivity of the X-ray cluster observable 
$X\sim d_a^{-3} \om^{-2}$ to the cosmological 
parameters $p=\{\om,w_0,w_a\}$ for a flat universe 
are encoded in the derivatives plotted here. 
The larger the absolute magnitude of the derivative at a particular 
redshift, the more constraining the observations there. 
}
\label{fig:sens} 
\end{center} 
\end{figure}

Figure~\ref{fig:sens} shows the run of unmarginalized cosmological 
parameter sensitivity with redshift.  Clearly the matter density $\om$ 
is the best determined parameter, and furthermore its sensitivity 
$\partial X/\partial\om$ 
has substantial redshift dependence, so cluster measurements over a wide 
redshift range will likely improve constraints on $\om$.  Sensitivity to 
dark energy parameters is much reduced, in particular to the time variation 
$w_a$.  Note that at $z>1$, the sensitivity curves of $w_0$ and $w_a$ are 
not parallel so a deeper survey can break the degeneracy between them 
(though of course precision observations of high redshift X-ray clusters 
are more difficult). 

To take into account the degeneracies between cosmological parameters, 
we employ a Fisher analysis involving 10\% determinations of $X$ in 
each bin of 0.1 in redshift over the range $z=0.1-0.9$ or $z=0.1-1.7$.  
One can think of such precision naively as corresponding to better than 
3.3\% measurements of distance; 
we leave until \S\ref{sec:bias} the issue of what is the contribution 
of statistics vs.\ systematics to this number.  This 10\% value is 
purely illustrative; we consider later in this section what level of 
accuracy would be required to achieve substantial constraints on the 
dark energy parameters; in the absence of priors, errors and likelihood 
contours scale with the value adopted for the uncertainty.  We also 
emphasize that these uncertainties 
are not measurement errors; they represent the amalgamated uncertainties 
(systematic plus statistical, with systematics likely to dominate) on 
the quantities entering the cosmological expression as defined by 
Eq.~(\ref{eq:fxd}).

\begin{figure}[!hbt]
\begin{center} 
\psfig{file=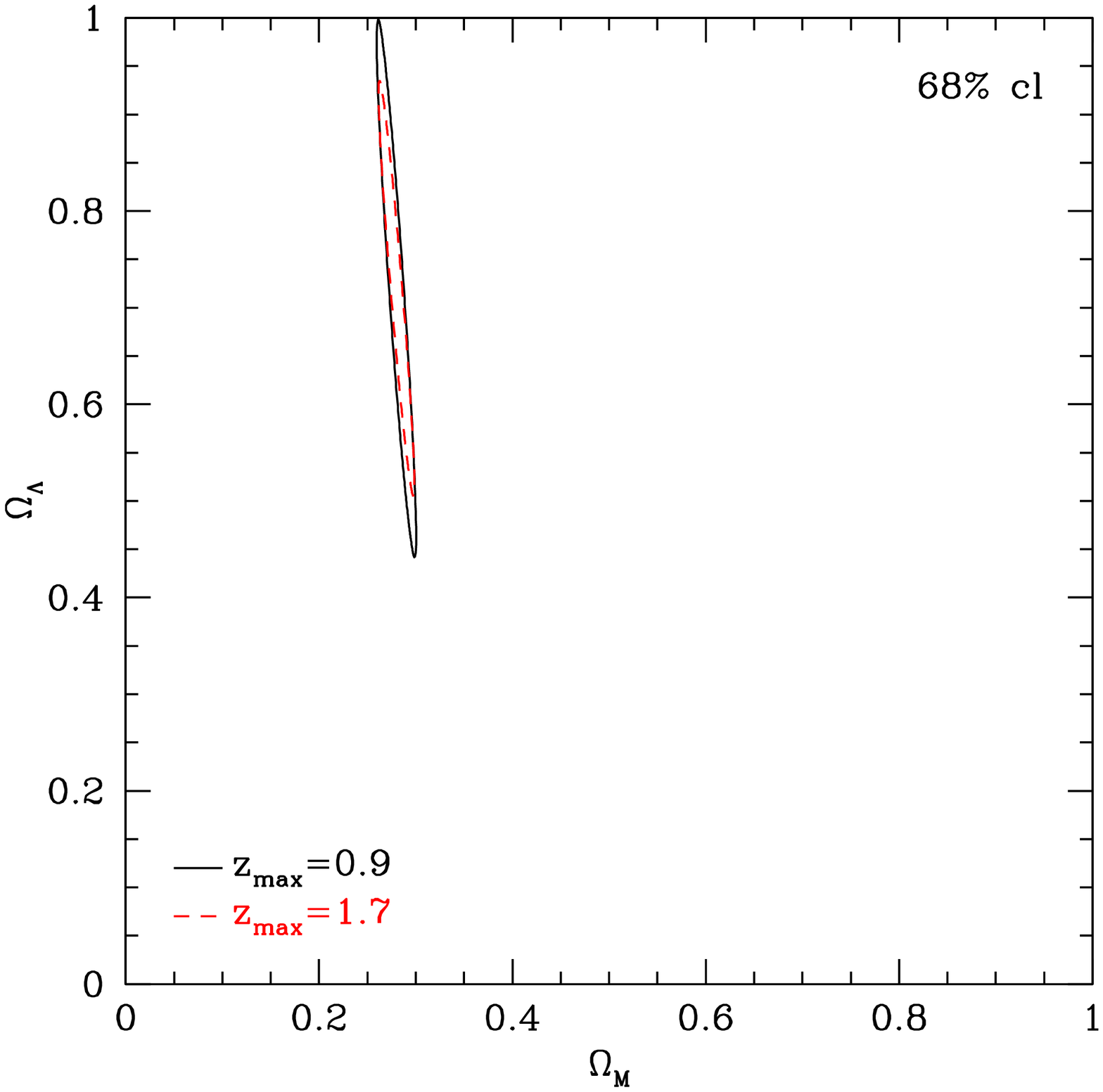,width=3.4in} 
\psfig{file=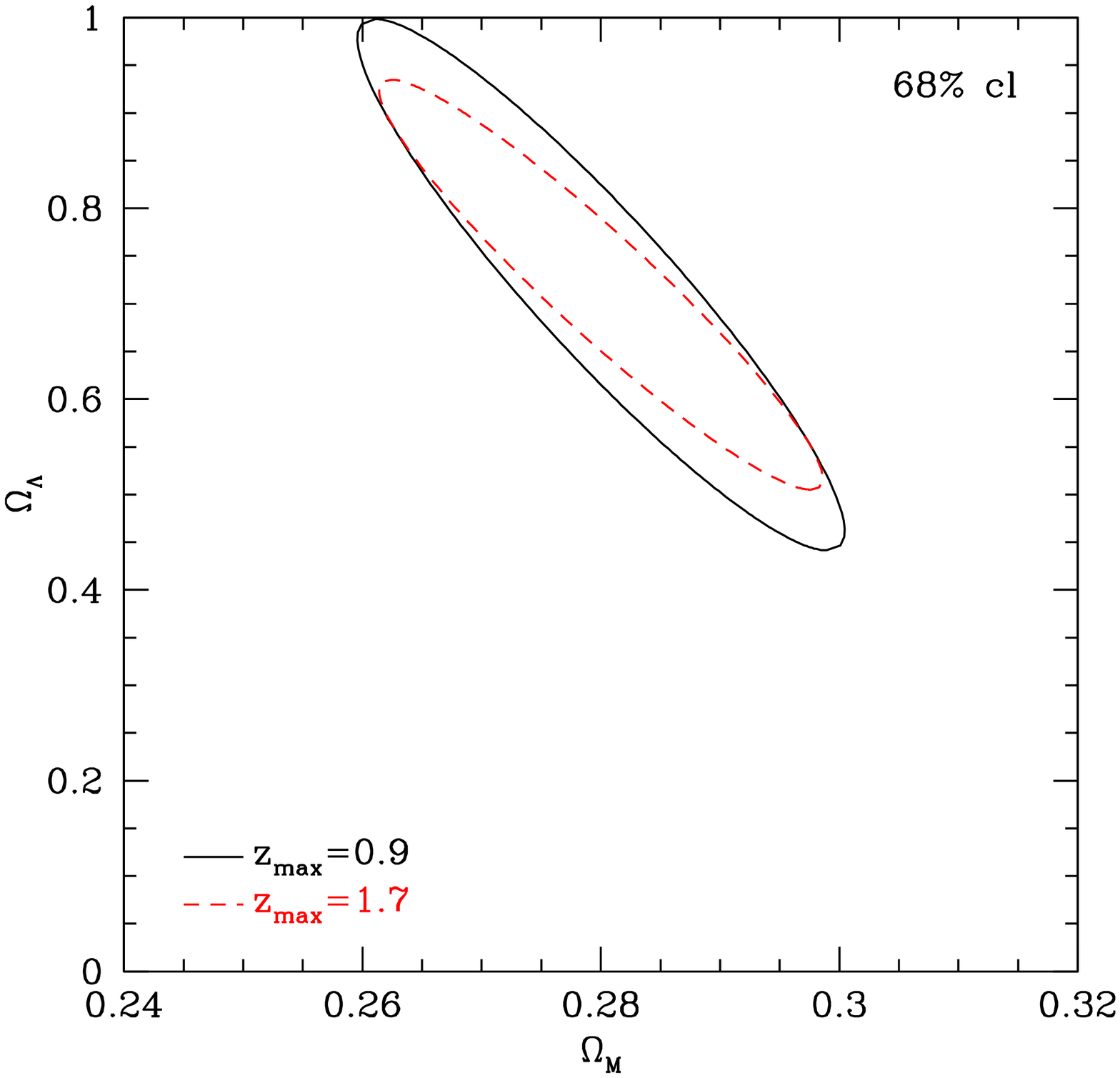,width=3.4in} 
\caption{X-ray data constraints in the matter density vs.\ 
cosmological constant energy density plane act primarily to determine 
the matter density.  The $1\sigma$ joint confidence contours show 
the results from simulated X-ray cluster measurements out to $z=0.9$ 
(e.g.\ current Chandra observations) or $z=1.7$ (e.g.\ next generation 
Con-X observations), with 10\% determinations of the parameter 
combination $X=d_a^{-3}\om^{-2}$ in each redshift bin of width 0.1 
(under the idealized case of no systematic biases and other parameters 
fixed).  The bottom panel zooms in on a region of the top panel. 
}
\label{fig:omlam} 
\end{center} 
\end{figure}

First, we look at the constraints on matter density and cosmological 
constant in a possibly nonflat universe.  The contours 
in Fig.~\ref{fig:omlam} (holding fixed $w=-1$) are strongly 
aligned to give tight constraints on $\om$, with much weaker bounds 
on $\Omega_\Lambda$.  Indeed, $\om$ is 10-15 times better determined 
than $\Omega_\Lambda$.  
The orientation of the contour is in good agreement with similar 
plots in \cite{ettori,allen04,rapetti}.  The extent of the contours will vary 
with different inputs for the redshift range and systematics level.  
Note we have also held the amplitude $A$ fixed (including the depletion 
bias and the baryon-gas offset, as well as 
fixing $\omb$, all to be discussed in \S\ref{sec:bias}), 
making this overidealized.  Even so, a systematics level of $<2.5\%$ 
(roughly equivalent to 0.8\% distance measurements) is required to 
determine $\Omega_\Lambda$ to 5\% from a survey to $\zm=1.7$.

\begin{figure}[!hbt]
\begin{center} 
\psfig{file=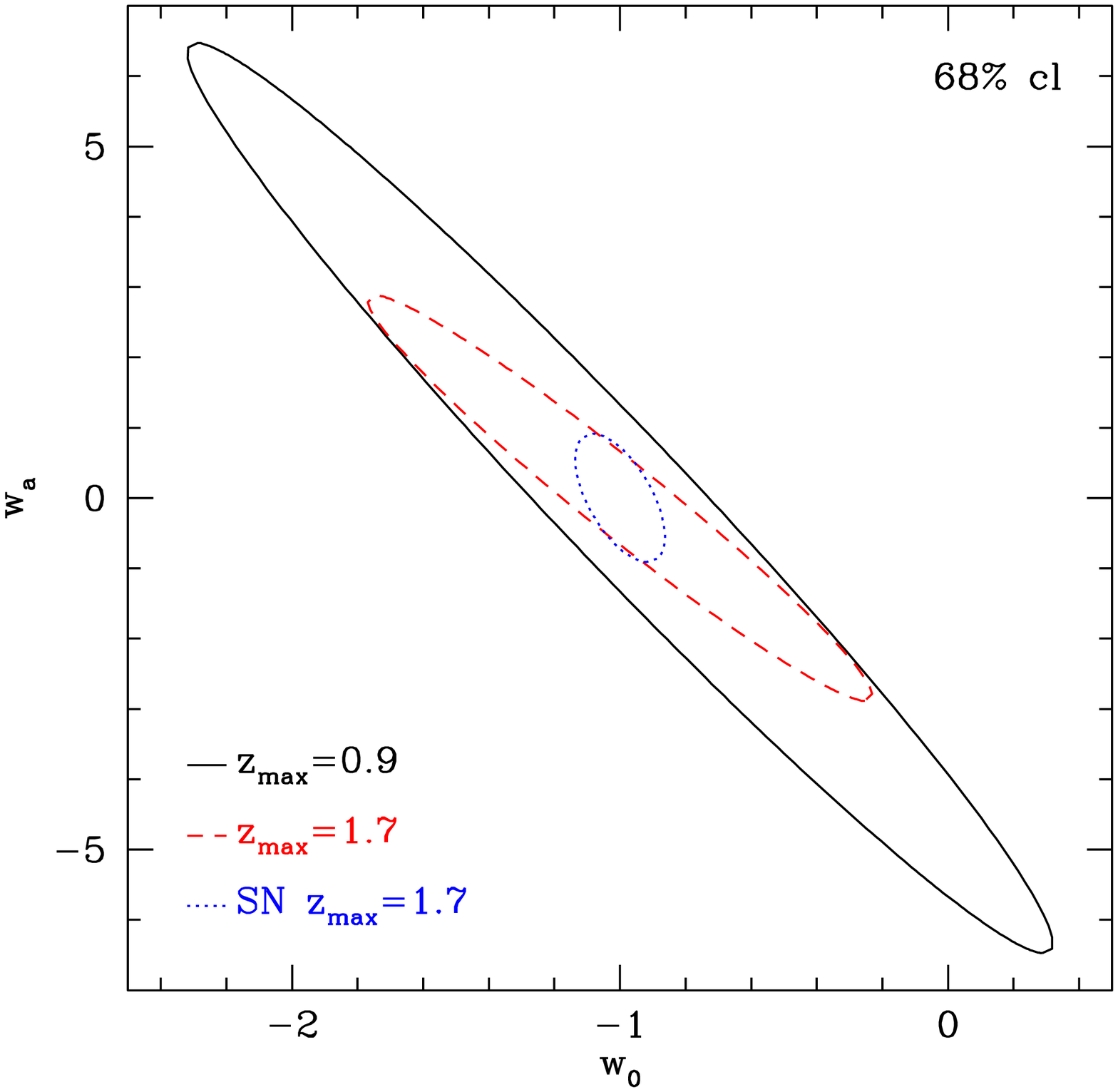,width=3.4in} 
\caption{X-ray data constraints, simulated as in Fig.~\ref{fig:omlam} 
(i.e.\ with idealized assumptions), provide only weak limits on the 
dark energy equation of state properties.  
Even a $\zm=1.7$ X-ray survey (e.g.\ Con-X depth) 
does not reveal dark energy properties at a level near 
the constraints expected from supernova distance measurements out to 
$z=1.7$ (e.g.\ SNAP quality, including systematics). 
}
\label{fig:wowa} 
\end{center} 
\end{figure}

Using X-ray cluster gas mass as a probe of dark energy cosmology, 
where the equation of state is fit by $w_0$ and $w_a$ but restricting to 
a flat universe, we find that the constraints are less promising.  If we 
restrict to a (physically unjustified) constant equation of state, 
then the X-ray data in the idealized case is oriented in basically the 
same direction as the CMB or weak lensing probes, with good orthogonality 
with supernova distance data.  Allowing for the possibility of time 
variation of the dark energy equation of state, however, the 
insensitivity we saw in Fig.~\ref{fig:sens}  
means that the X-ray technique is poor at uncovering the nature of 
dark energy.  For cluster data extending out to $z=1.7$, even in the 
idealized case of fixed amplitude $A$, the constraints on dark energy 
are still 
$\sigma(w_0)\ga 0.5$ and $\sigma(w_a)\ga 2$; see 
Figure~\ref{fig:wowa}.  (If one insists on considering a constant 
equation of state, the uncertainties can be found by taking a cut 
across the confidence contours, holding fixed $w_a=0$.)

Combining X-ray cluster gas mass measurements with other probes of 
dark energy does not greatly improve the constraints. 
X-ray data to $\zm=1.7$ (XR) plus Planck 
CMB data on the distance to the last scattering surface still do not 
constrain $w_0$ to better than 0.4 or $w_a$ to better than 1.  Adding XR 
to a baseline of supernovae (SN) to $z=1.7$ plus CMB plus weak lensing 
shear power spectrum (WL) of a 1000 deg$^2$ space survey, 
tightens the area of the $w_0$-$w_a$ contour by only 2\%.  Substituting 
XR instead of SN blows up the area by a 
factor 3.  Furthermore, such a combination would possess no purely 
geometric probe of cosmology (independent of nonlinear structure 
formation and gas hydrodynamics). 

To ensure that these conclusions are robust, we have carried out several 
further investigations.  We 
have checked that the exact distribution of clusters within 
a given redshift range does not significantly affect the results.  This 
is the case for other distance probes as well, where the optimum 
redshift distribution gives errors within a few percent of a flat 
distribution or a smoothly sculpted one \cite{huttur}.  For example, 
changing the uniform distribution to a Gaussian centered on $z_{\rm max}/2$ 
with standard deviation $\zm/4$, 
our cosmological parameter estimations change by less than 4\% (10\%) 
for $\zm=0.9$ (1.7). 

From the sensitivity plot, Fig.~\ref{fig:sens}, we had conjectured 
that extending the redshift range beyond $z\approx1$ would continue 
to provide improved constraints (even in the $\Lambda$ fiducial model).  
By keeping the same 
number of independent data points as for the $\zm=1.7$ survey, but 
varying the survey depth $\zm$, we confirm that the depth, not the 
greater number of statistical data bins as the redshift range increases, 
is predominantly responsible for improved cosmology estimation. 
The gain from depth saturates around $\zm\approx1.7$, e.g.\ the case 
with $\zm=2$ only gives a further 5\% (9\%) improvement on $w_0$ ($w_a$). 

If we think that we can neglect all systematics and continue to 
improve constraints purely with more statistics without a floor, 
then to achieve estimation of $w_0$ to better than 0.1 or $w_a$ to 
better than 0.5, we require measurements of $X$ to 2\% (roughly 
0.67\% distance accuracy) -- although this still relies on an 
idealized analysis, without the necessary marginalizations and possible 
biases discussed in the next section.  This appears to be challenging 
for the required level of astrophysical understanding and telescope time 
(\cite{rapetti} estimates that applying this technique to 250-500 
clusters will require 10\% of all Con-X observing time in the first 
five years).

\section{Systematics and Biases} \label{sec:bias} 

The analysis up to this point of X-ray data has been idealized by ignoring 
possible systematic effects from the non-constancy of $A$ (including 
$\fg$) in Eq.~(\ref{eq:fxd}).  
That is, even if the precision of a sample at a given redshift is 
extremely good, a trend with redshift acts as a bias on 
cosmological parameter estimation.  Therefore, any uncertainty on such 
a trend acts as a systematic uncertainty of the results. 

As mentioned in the Appendix, one possible uncertainty concerns the 
recognized correction of the gas mass fraction by the additional 
factor \cite{allen04} 
\beq
\fg\sim [b/(1+0.19\sqrt{h})]\,\fg^{\rm ideal}, \label{eq:fbias}
\eeq 
representing the effects of depletion bias $b$ of baryons leaking out of 
clusters during formation and relaxation \cite{eke98}, and the correction 
for cluster baryons present in galaxies rather than intracluster gas 
\cite{white93}.  Recall that $\fg$ enters squared into $X$.  
This, and other effects outlined in the Appendix that characterize the 
gas properties, can all be examined 
through the amplitude factor $A$.  If this parameter is not assumed 
a priori fixed, it must be marginalized over (similar to the case 
of the absolute magnitude in supernova cosmology); since we have 
ignored this up to now, 
cosmological constraints from X-ray data derived in the previous 
section are overoptimistic.  The 
uncertainties in $A$, as well as possible drifts with redshift in its 
value, e.g.\ from evolving cluster or cluster environment properties, 
generate systematic floors beyond which statistical improvement is cut off. 

Taking proper account of the marginalization over $A$ (which in 
principle includes not only the astrophysical factors, but the baryon 
density $\omb$) will act to blow up the uncertainties on the 
cosmological parameters, especially $\om$ and $w_a$.  If $A$ is not 
limited a priori, then to achieve cosmology estimation with $\sigma(w_a)<1$ 
we find measurement of thousands of times as many clusters are required 
than when $A$ is assumed completely known.  For this many clusters to 
contribute usefully, systematic uncertainties must be bounded below 
the 0.2\% level.  

First consider the case where $A$ is a random variable, independent 
of redshift. 
If we understand the gas hydrodynamics and other astrophysical 
uncertainties, using the full range of X-ray and other cluster 
measurements, we can constrain $A$ to some precision with a prior. 
In the previous section we found that in order to obtain significant 
constraints on dark energy, e.g.\ knowing $w_a$ to within 0.5, we required 
a systematics level of less than 2.6\%.  Given that level, but now 
removing the idealization of perfectly known $A$, calculations indicate 
that predicting 
$A$ to 3\% (10\%) degrades the estimation of $\om$ by 30\% (a factor 3). 

Now consider not a statistical uncertainty, but a systematic 
trend in redshift.  Again, the array of precise X-ray measurements 
can attempt to correct for this by fitting the evolution in $A$ 
with extra parameters -- {\it if\/} 
we know the functional form (sometimes called self-calibration).  Suppose 
we guess $A=A_0(1+z)^\alpha$.  Even with $A_0$ fixed, fitting for $\alpha$ 
blows up beyond use the cosmological parameter determination, i.e.\ 
pure self-calibration fails.  We 
therefore require priors for $A_0$ and $\alpha$, i.e.\ we must understand 
the astrophysical properties of clusters sufficiently well to limit these 
parameters.  We find that $A_0$ basically is degenerate with $\om$ 
(cf.\ Eq.~\ref{eq:fxd}) and 
$\alpha$ has strong degeneracy with $w_a$. 

If we misestimate the values of the calibration parameters then the 
residual systematics bias the cosmology.  For example, mistaking $A_0$ 
leads to a bias $\delta\om/ \sigma(\om)=0.73\,(\delta A_0/0.1)$.  If 
we require keeping the offset below $0.46\sigma$ (so the risk, the 
square root of the quadratic sum of the statistical and systematic 
errors, is degraded by less than 10\%), then $A_0$ must be known 
to 0.06.  Such a redshift independent shift (as opposed to the 
previous dispersion) does not have a strong 
effect on the cosmological model. 

\begin{figure}[!hbt]
\begin{center} 
\psfig{file=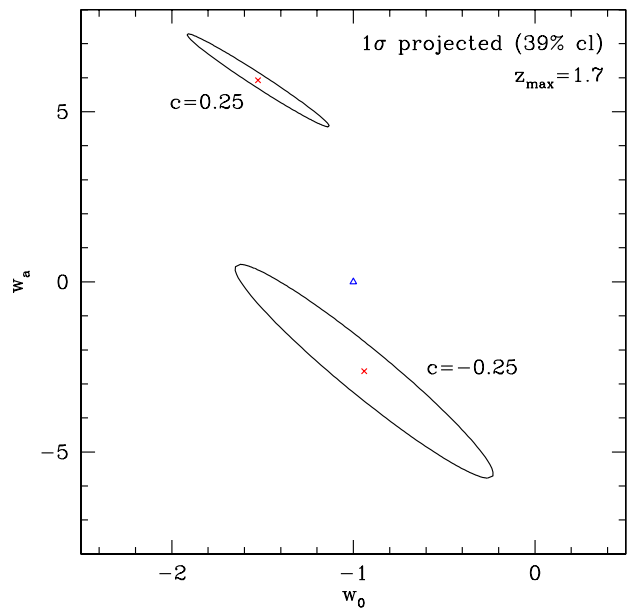,width=3.4in} 
\caption{Systematic uncertainties in X-ray cluster properties can 
strongly bias the cosmological parameter estimation.  A mild evolution 
in the baryon depletion or galaxy contribution factors (see 
Eq.~\ref{eq:fbias}), for example, going with redshift as $(1+z)^c$ can 
shift the derived cosmology (shown in each case by x's) by several 
standard deviations from the true cosmology (given by the open triangle). 
}
\label{fig:biasb} 
\end{center} 
\end{figure}

Residual uncertainty in the redshift dependence, in the simplest 
case the value of the power law index $\alpha$, is more significant. 
Perhaps the simplest scenario is assuming, say, the depletion 
factor $b$ in Eq.~(\ref{eq:fbias}) is universal, when in fact it 
has some evolution $\sim(1+z)^c$ (hence $\alpha=2c$). 
Figure~\ref{fig:biasb} shows the rather dramatic results of bias in 
the cosmological parameters generated by an evolution as slow as $c=\pm0.25$. 
(We plot $1\sigma$ projected, or 39\% confidence level, contours so 
the biases can read off directly from projection to the axes.)  
Such misestimated evolution can bias the true cosmology by $3\sigma$.

To ensure $w_a$ is biased less than 0.5 off the true value, one 
requires $\delta c<0.03$.  That is, an error in understanding 
redshift evolution of $\fg$ of $(1+z)^{0.03}$ will invalidate 
the cosmology estimation.  This is further complicated by 
the fact that measuring the baryon depletion $b$ from observations 
is itself entangled with cosmology.  Knowing $c$ to 0.03 implies 
an understanding of baryons in clusters, at low and high redshift, 
such that a drift of 3\% in $b$ from $z=0$ to 1.7 can be ruled out. 
Furthermore, gas properties discussed in the Appendix, such as 
clumpiness, and temperature and pressure profiles, multiply $\fg$ 
in the amplitude factor $A$ and so the entire product must be 
subject to this tight systematics control.  Note that self-calibration 
does not immunize against such systematic errors unless the adopted 
evolutionary function is exact -- i.e.\ one must know the form of the 
variation beforehand for self-calibration to work. 

In addition to physical evolution in cluster properties, systematic 
uncertainties can also arise from 
population drift, e.g.\ if there are multiple populations with different 
baryon depletion factors, say, and population $i$ is more prevalent 
at low redshift and population $j$ dominates at high redshift. 
We have seen that such redshift dependent systematics have a strong 
influence on $w_a$, while conversely the inherent sensitivity of the 
X-ray cluster baryon/gas mass fraction method to $w_a$ is low (see 
Fig.~\ref{fig:sens}), so it appears quite challenging for even next 
generation X-ray cluster distances to provide 
robust constraints on dark energy properties. 

To estimate in detail where the systematic floor lies for this technique 
is beyond the scope of this paper, and will require substantial 
understanding of cluster mass distributions and gas properties. 
Some of the possible systematics identified are listed in the Appendix. 
We have shown that in order to achieve determination $\sigma(w_a)<0.5$ 
we must limit the redshift independent systematic uncertainty to below 
0.7\% on both the gas mass fraction and the distance determination.  
If such a systematic floor could be reached, then matching this error level 
with statistics would require a sample of more than 
\beq 
N=3400\left(\frac{D}{0.1}\right)^2\ {\rm clusters} 
\eeq 
out to $\zm=1.7$, to statistically reduce an individual cluster 
dispersion of $D$ down to a level $0.7\%=D/\sqrt{N}$. 
For the combination of 
the X-ray technique with Planck CMB data, achieving $\sigma(w_a)<0.5$ 
requires 1\% systematics and 1600 tightly characterized clusters. 
Note that proposed large 
programs for Con-X may measure a few hundred clusters \cite{rapetti}. 

As we saw, the redshift {\it dependent\/} systematic uncertainty has 
more severe effects.  We must know not only the functional form of 
any evolution or population drift in cluster properties, but fit 
the parameters of the evolution to high accuracy.  For example, a 
residual drift more rapid than $(1+z)^{0.03}$ will bias the time 
variation of the dark energy equation of state, $w_a$, by more than 0.5.

\section{Conclusion} \label{sec:concl}

We have considered the possibility of current and next generation 
observations of X-ray clusters as cosmological distance probes through 
the baryon/gas mass fraction method.  In current and future usage, assuming 
a cosmological constant universe, this technique can provide tight 
constraints on the matter density.  However, even with the next generation 
Con-X and XEUS observatories, we find that it will be a considerable 
challenge for such 
X-ray cluster measurements to provide a precision dark energy probe.  

As with any probe, we cannot ignore possible systematic 
uncertainties.  While many have been identified (see the Appendix for 
a brief discussion), the level of systematics control possible is 
currently not well determined.  Understanding of cosmic geometry, 
the nonlinear mass distribution, and gas hydrodynamics are simultaneously 
required.  The amplitude factor $A$ must be fit for, and redshift 
dependent uncertainties from evolution or population drift are especially 
degenerate with the dark energy equation of state time variation. 
Self-calibration only works if the functional form of the evolution is 
known and if the redshift scaling can be calculated to within $(1+z)^{0.03}$. 

X-ray cluster studies from simulations, theory, and observations over 
the next decade will address what is a 
realistic level of systematic uncertainties, both as a floor on precision 
at a given redshift and as a redshift dependent bias.  We emphasize that 
this paper does not claim what level should be adopted but provides a 
general investigation of what level will impact cosmological parameter 
determination.  Even if the X-ray gas mass fraction technique turns out 
not to be significant as a dark energy probe, the X-ray data is likely to 
greatly extend and deepen our knowledge of cluster properties and evolution.

\ack 

I thank Jochen Weller and especially the referee for helpful comments. 
This work has been supported in part by the Director, Office of Science,
Department of Energy under grant DE-AC02-05CH11231. 

\appendix 

\section*{Appendix: Review of Method} 

\setcounter{section}{1}

The gas mass fraction of a galaxy cluster, traced through X-ray 
measurements, is neither precisely a baryon fraction measurement 
nor a distance indicator.  It is useful to briefly, if basically, 
review the quantities that enter, and under what 
assumptions.  We follow the approach of \cite{sasaki}; also see 
\cite{allen04,weller}.  As with any cosmological probe, systematic 
controls are necessary for each assumption.  

The measured X-ray flux $F_X$ is related to the intrinsic cluster X-ray 
luminosity $L_X$ and the cosmological luminosity distance $d_l$ to the 
cluster by 
\beq 
F_X=\frac{L_X}{4\pi d_l^2}. 
\eeq 
The luminosity distance is $H_0 d_l=(1+z)(1-\Omega_T)^{-1/2}\sinh 
[(1-\Omega_T)^{1/2}\int_0^z dz'/[H(z')/H_0]]$, where $\Omega_T=\om+\Omega_w$ 
is the total density in units of the critical density, and the Hubble 
parameter $[H(z)/H_0]^2=\om(1+z)^3+\Omega_w(1+z)^{3(1+w_0+w_a)}  
e^{-3w_a z/(1+z)}+(1-\Omega_T)(1+z)^2$. 
To use the cluster as a cosmological distance probe we need 
to understand the astrophysical and cosmological dependence of the cluster 
luminosity.  This adds requirements on understanding ``mass+gas'', i.e.\ 
nonlinear structure formation and gas hydrodynamics, to the geometric 
distance.  (See the cosmology probe classifications discussed in 
\cite{lindark}.)  For an analogous situation of astrophysical dependence 
of luminosity when using Type Ia supernovae as distance probes, and 
methods of treatment, see, e.g., \cite{coping}. 

The X-ray luminosity (taking into account the main dependencies) is 
\beq 
L_X\sim n_e^2T^{1/2}d_a^3, \label{eq:lum}
\eeq 
where $n_e$ is the free electron density, $T$ is the gas temperature, 
and $d_a=d_l/(1+z)^2$ is the angular diameter distance to redshift $z$. 
There are several well known issues.  The electron density enters squared, 
which makes 
cluster core detection easier but introduces sensitivity to substructure 
or clumpiness $\langle n_e^2\rangle/\langle n_e\rangle^2$.  The gas is 
assumed to be at an equilibrium temperature.  High resolution and 
spectral observations can test both of these to some extent. 
The distance enters through the volume the gas is occupying, given 
an observed angular size, which uses an assumption of spheroidal 
symmetry to deproject the two dimensional X-ray 
measurements to the three dimensional gas distribution.  The distribution 
may be spatially complex, with filaments and voids, but again this can 
be checked to some extent through observations.  If the cluster gas 
is relaxed to a smooth, ellipsoidal shape, one might think the average 
over orientations of many 
clusters would still give the correct luminosity.  However, there is a 
bias in that clusters prolate along the line of sight have a 
higher projected flux and hence 
are more easily detected; this amplification is proportional to the 
quadrupole moment of the distribution.  If the cluster flux is near 
the detection threshold then the quadrupole distribution does not 
average to $\langle Q\rangle=0$ but is offset due to selection bias. 

The electron density at some radius $R$ is proportional to the gas mass 
within a shell at that radius, $dM_g\sim n_e R^2 dR$.  To relate the 
radius to an astrophysical property of the cluster requires an 
assumption such as hydrostatic equilibrium, 
\beq 
G\mu m_H\frac{M_{tot}(<R)}{R}=-kT\,
\left(\frac{d\ln n_e}{d\ln R}+\frac{d\ln T}{d\ln R}\right), \label{eq:hydro}
\eeq
where $\mu m_H$ is the average mass per baryon.  Under the further 
assumption that the right hand side is cosmology independent, i.e.\ 
depending only on the local cluster properties, we find that $M_{tot}\sim 
R\sim d_a$.  It is not clear that either of these assumptions is 
guaranteed.  For example, if we choose $R$ to be at a fixed multiple 
of either the critical or background density at the cluster redshift, 
e.g.\ 2500 times ($r_{2500}$), then the mass depends on the cosmology 
and matter density. 
This is the standard problem with defining the mass of a cluster, since 
that is not necessarily a unique quantity.  
Furthermore, even if the 
cluster profile is universal, the concentration (related to the 
overdensity relative to the scale radius \cite{nfw97}) depends on 
cosmology \cite{bartelmann}, affecting 
the right hand side of Eq.~(\ref{eq:hydro}).  
As well, the temperature 
is possibly the greatest source of observational uncertainty and must 
be measured at the same radius as used for the cosmology; it will also 
be affected by non-uniformity or flux dependence of the spectral response 
of the X-ray instrumentation. 

Carrying forward regardless, we obtain from translating $n_e$ into $M_g$ 
and $R$ into $M_{tot}$ that 
\beq 
F_X\sim R^{-1}d_a^{-2}\left(\frac{dM_g}{dM_{tot}}\right)^2 
\sim d_a^{-3}\left(\frac{dM_g}{dM_{tot}}\right)^2 
\sim d_a^{-3}\left(\frac{\Omega_b}{\Omega_m}\right)^2. \label{eq:flux} 
\eeq 
The last expression converts from gas quantities to baryon quantities. 
A major 
assumption in the use of Eq.~(\ref{eq:flux}) is that the gas mass fraction 
$M_g/M_{tot}$ measured for any cluster is equal to a constant quantity 
$f_{\rm gas}$, and that furthermore this is the universal baryon to matter 
ratio $\omb/\om$.  In fact, since baryons are present in cluster galaxies 
as well as the intracluster (X-ray emitting) gas, and baryons are also 
lost during cluster formation, one must make a correction as in 
Eq.~(\ref{eq:fbias}).  

We thus end up with the central cosmological dependence shown in 
Eq.~(\ref{eq:fxd}), if the assumptions hold or can be corrected for. 
All the factors of non-universality of gas clumpiness, gas pressure 
and temperature profiles, and the gas mass fraction $\fg$, multiply 
together to form the amplitude factor $A$ for the proportionality in 
Eq.~(\ref{eq:flux}).  Any cosmology dependence in the gas parameters 
in $A$ alters the cosmology deduced for the true cosmological 
quantities of the distance $d_a$ and matter and baryon densities 
$\om$ and $\omb$. 

An additional point that may seem straightforward, but does have the 
potential to be overlooked, concerns a nonlinear transformation of 
the analysis variables.  We use $X$, as the quantity most closely related 
to observables.  If phrased in terms of $\fg$ or $d_a$, the transformation 
is nonlinear, and Gaussian errors in the observables will become 
non-Gaussian.  Put most simply, 
$\langle d_a\rangle\sim\langle F_X^{-1/3}\rangle\ne \langle 
F_X\rangle^{-1/3}$.  As is known from the analysis of supernovae, 
when transforming fluxes into magnitudes (logarithms of flux), such 
non-Gaussianities, if overlooked, can lead to bias in the derived 
cosmology, as emphasized generally by \cite{holzlin}.  This can be 
nontrivial: for example, a Gaussian dispersion $\sigma_X/X=0.02z$ that 
is analyzed in terms of $\fg$ with Gaussian errors will turn a true 
cosmological constant cosmology into one that looks like evolving dark 
energy with $w_a\approx-0.5$.

\end{document}